\documentclass[draft]{agujournal2019}
\usepackage{url} 
\usepackage{lineno}
\usepackage{subfigure}
\usepackage[inline]{trackchanges} 
\usepackage{soul}
\draftfalse

\journalname{arXiv}

\begin{document}

%
%


\title{Predictive modeling of altitude resolved greenline airglow emission (557.7 nm) in the MLT region}

%
%




\authors{{Dayakrishna Nailwal}\affil{1}, M V Sunil Krishna\affil{1,2}, Alok Kumar Ranjan\affil{1,3},\\ D. Pallamraju\affil{3}}


\affiliation{1}{Department of Physics, Indian Institute of Technology Roorkee,  \\ 
Roorkee-247667, Uttarakhand, India}
\vspace{3mm}
\affiliation{2}{Centre for Space Science and Technology, Indian Institute of Technology Roorkee, \\Roorkee-247667, Uttarakhand, India}
\vspace{3mm}
\affiliation{3}{Space and Atmospheric Sciences Division, Physical Research Laboratory, Ahmedabad, India}




\correspondingauthor{Dr. MV Sunil Krishna}{mv.sunilkrishna@ph.iitr.ac.in}




\begin{keypoints}
\item The machine learning approach can be a promising method to forecast space weather modulation of Earth’s atmosphere.
\item A machine learning model has been developed to predict the volume emission rate of greenline emissions using the measurements of WINDII.
\item The model’s predictions match very closely with the actual satellite-based measurements and those estimated from the GLOW model.
\item The model effectively captures the variability of atomic oxygen green line emission with respect to a variety of geophysical parameters.
\end{keypoints}

%
%

%
%


\begin{abstract}
\justifying
Atomic Oxygen (O) is a critical and highly reactive chemical species responsible for key physical and chemical processes in the mesosphere and lower thermosphere (MLT). Understanding the variability of atomic oxygen is crucial for comprehending the intricate processes and dynamics in Earth’s upper atmosphere. The airglow emission resulting from the atomic oxygen at 557.7 nm is a highly sensitive tracer for remotely sensing the MLT processes. The limited availability of global direct observations of these emissions limits the understanding of space weather modulation of the upper atmosphere. This study introduces a novel and highly accurate method based on machine learning to estimate the atomic oxygen green line airglow emission in the upper atmosphere. By incorporating the geomagnetic and solar forcing parameters, this research offers a fresh perspective on studying Earth’s upper atmosphere. An atomic oxygen emission machine learning model (AOEMLM) is developed using 13 years of Wind Imaging Interferometer (WINDII) measurements. All primary solar and geomagnetic parameters that influence the chemistry and energy balance in the upper atmosphere of Earth are considered in the development of AOEMLM. The training and testing of AOEMLM is conducted using the WINDII dataset and the random forest ML algorithm. AOEMLM is validated using the WINDII measurements from the years 2001-2003. The model successfully captures the variability in atomic oxygen (O) greenline vertical profiles with good accuracy. The predictions of AOEMLM are verified using WINDII and ICON-MIGHTI measurements and with the existing chemistry-based GLOW model. Results show that the AOEMLM broadly agrees with the satellites and photometer observations.
\end{abstract}

\section*{Plain language summary}
\justifying
The mesosphere and lower thermosphere are chemically active and dynamic regions of the Earth's upper atmosphere. This region is the first point of contact when radiation or energetic particles enter Earth's proximal space. Atomic oxygen participates in many physical, chemical, and dynamic processes. An optical emission resulting from atomic oxygen at a wavelength of 557.7 nm serves as a very important tracer for remotely understanding the state and structure of the MLT region. In this study, a machine learning model known as AOEMLM based on thirteen years of WINDII observations is reported. The model is developed using satellite data combined with several solar forcing parameters, such as the F10.7 cm solar flux and the sunspot number, along with the solar zenith angle (SZA) and local time (LT). The F10.7 cm solar flux and sunspot number are indicators of solar activity, while SZA and LT indicate the position of the Sun in relation to the local coordinates. These inputs can be used as proxies for the incident solar radiation in day and night conditions. Recent advances in machine learning and new algorithms provide promising opportunities for modeling and to better understand the effect of space weather. The model AOEMLM is designed to estimate the vertical profile of green line volume emission rate (VER) and trained using the atomic oxygen green line emission (557.7 nm) dataset measured by the WINDII instrument onboard Upper Atmosphere Research Satellite (UARS) from 1991 to 2000. Its predictions were then validated using the WINDII dataset from 2001 to 2003. Additionally, the prediction capability and performance of the AOEMLM were tested using the ICON-MIGHTI and ground-based photometer datasets. The results indicate that the AOEMLM successfully captures the variability in atomic oxygen (O) greenline vertical profiles and shows good agreement with the satellite and photometer measurements.

\section{Introduction}
\justifying
The upper atmosphere, particularly the thermosphere, plays an essential role in Earth's atmospheric system, extending from approximately 85 km to 600 km above the Earth's surface. It is directly exposed to solar radiation, energetic particles, and space weather effects, leading to significantly high temperatures and higher levels of ionizing radiation compared to the lower atmospheric layers, such as the troposphere and stratosphere. During extreme space weather events, such as geomagnetic storms, solar flares, coronal mass ejections (CME), and high-speed solar wind, the temperature and ionizing radiation in the upper atmosphere are enhanced significantly compared to the normal conditions. These space weather events are accompanied by a massive amount of energy in the form of ionizing radiation and particle precipitation into the upper atmosphere. The deposition of excessive energy in the upper atmosphere, especially in the MLT region, and the resulting increase in ionizing radiation trigger various physical, chemical, and dynamical processes that change the atmospheric chemistry, which leads to the creation and depletion of various reactive species, including ozone (O$_3$), atomic oxygen (O), hydroxyl radicals (OH), nitrogen oxides (NOx), and other atmospheric constituents \cite{turunen2009impact,mansilla2011some}. Atomic oxygen (O) is one of the important and dominant reactive species present in the upper atmosphere of Earth and plays a vital role in the chemistry and dynamics of MLT region \cite{ward1993role}.
Its abundance and behavior in the MLT region are influenced by various factors such as solar activity, atmospheric composition, and dynamics. These variations have significant implications for climate variability and space exploration.
Atomic oxygen (O) reacts with several atmospheric species, such as ozone (O$_3$) and nitrogen (N$_2$) to produce a variety of compounds, which have significant consequences on the energy balance and circulation pattern of the upper atmosphere such as nitric oxide (NO), which plays a key role in the energy budget and dynamics of the MLT region \cite{mlynczak1993detailed, mlynczak2013atomic}.

In addition to its chemical effects, atomic oxygen also has a significant impact on the technology that we use in our daily lives. It can have a drastic impact on satellite operations, as it offers a drag on the satellite motion. It can cause degradation of materials used in spacecraft and satellites, especially in low earth orbit (LEO) \cite{reddy1995effect} due to its high reactivity. Apart from this, when charged particles from the Sun enter the Earth's upper atmosphere, they collide with atoms and molecules in the upper atmosphere, including atomic oxygen. These collisions excite the atomic oxygen, causing it to emit light in the form of aurora \cite{savigny2017airglow, singh19961}.

The production of atomic oxygen (O) in the mesosphere and thermosphere goes through a series of complex chemical reactions that involve the interaction of solar energetic radiation (EUV and UV) with the Earth's atmosphere.
In the mesosphere and thermosphere, atomic oxygen is primarily produced through the photo-dissociation of molecular oxygen (O$_2$), molecular nitrogen (N$_2$, represented by reaction (2) \& (3)) and ozone (O$_3$) by solar ultraviolet (UV) ($\lambda$ $<$ 240 nm) \cite{brasseur1986recombination}  and extreme UV (EUV) radiation \cite{thorne1980importance}.

\begin{eqnarray}
     O_2 + h\nu (\lambda < 240 ~nm) &\rightarrow& O + O  \\
N_2 + h\nu &\rightarrow& N  +  N \\
N + O_2  &\rightarrow&  NO + O\\
O_3  + h\nu (\lambda < 240~ nm) &\rightarrow& O_2 + O
\end{eqnarray}

Additionally, atomic oxygen can also be produced by the ionization of molecular oxygen by high-energetic particles, such as cosmic rays and energetic electrons. This process is less common than dissociation by UV radiation but still contributes to the overall production of atomic oxygen in the mesosphere. The resulting nitrogen and oxygen atoms then combine to form nitric oxide (NO) and atomic oxygen (O), as represented by equation (3).
   
When an oxygen atom (in the ground state) gets excited either by collisional excitation with a high-energy electron, atom/molecule, or by various other chemical processes, it gets excited to a higher energy state. The higher energy state is unstable, and thus, the oxygen atom naturally tends to return to its ground state. As the atom returns to its ground state, it releases energy in the form of electromagnetic radiation. 
In the case of atomic oxygen, the energy released corresponds to the green line wavelength of light, well known as the atomic oxygen green line. The green line emission (557.7 nm) specifically results from the electronic transition of atomic oxygen from the second excited state O($^1$S) to its first excited state O($^1$D), as shown by the following equation.
\begin{equation}
    O(^1S)   \rightarrow  O(^1D) + h\nu ~(557.7 nm)
\end{equation}

\begin{figure}[h!]
\centering
\includegraphics[width=0.6\textwidth]{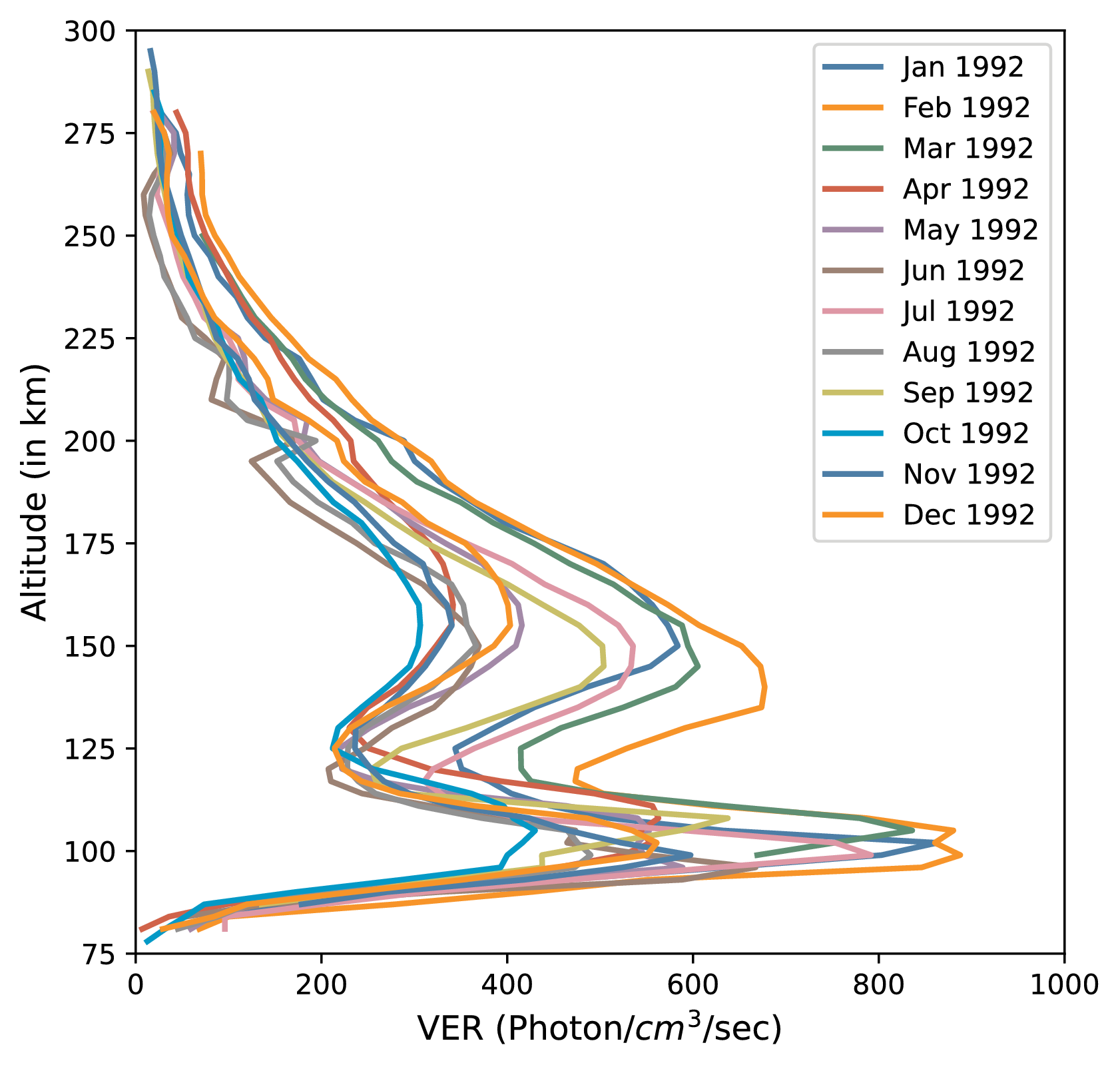}
\caption{The vertical daytime profiles of atomic oxygen green line volume emission rate during 1992.}
\label{fig1}
\end{figure}

The green line emission by atomic oxygen is a prominent feature of airglow and aurora, primarily occurs in the mesosphere and lower thermosphere, having its two intense peaks centered around 92-105 km (E-region of Ionosphere or upper mesosphere) and 140-180 km (F-region of Ionosphere or lower thermosphere) respectively \cite{zhang2005response, bag2017study}. Figure \ref{fig1} shows a few typical daytime reference profiles of atomic oxygen green line (557.7 nm) volume emission rates, randomly selected from the WINDII observations during the year 1992. As we can see from Figure \ref{fig1}, the upper mesospheric (92-105 km) and lower thermosphere peaks (140-180 km) are clearly visible.  It is to be noted that the production and loss processes of O($^1$S) green line emission are different in each production layer. In the lower thermospheric region, the primary source of excitation of green line emission is photoelectron impact, dissociative recombination of O$_2^{+}$, and the three-body recombination reactions. The detailed chemistry of atomic oxygen greenline emission is discussed in \cite{singh19961,barth1964three, thomas1981analyses,bag2017study}.

The green line emission by atomic oxygen (O) can be used as an important diagnostic tool for understanding the upper atmosphere, especially the MLT region. The measurement of the intensity and distribution of 557.7 nm green line emission can provide information about the temperature, density, composition, as well as the dynamics of various underlying atmospheric processes \cite{fauliot1997mean,ward1994correlations, garcia1985effect,zhu2015nighttime}. The earlier studies on airglow emission by atomic oxygen \cite{strutt1935light, deutsch2003long, reid2014seasonal} observed that the vertical profile of atomic oxygen volume emission rate (VER) shows variation with altitude and time and is influenced by a variety of processes, such as atmospheric dynamics, solar activity, and chemical reactions. The atomic oxygen green line (557.7 nm) VER is directly affected by the concentration of atomic oxygen, thermospheric temperature, and densities. These studies show that the atomic oxygen VER highly correlates with sunspot number and F10.7 solar flux. During extreme space weather conditions, when the densities of neutral and charged species in the upper atmosphere change dramatically due to a large amount of energetic particle precipitation, the enhancement in volume emission rate of atomic oxygen is observed. The variation in dayglow of atomic oxygen green line VER during a geomagnetic storm is also reported by \cite{bag2014effect, karan2018effect}. Several studies \cite{zhang2005response,russell2004atomic, russell2005atomic} also show that the VER by atomic oxygen is highly variable with solar zenith angle (SZA), latitude, local time, and altitude.

The atomic oxygen green line emission in the upper atmosphere exhibits seasonal variation \cite{pallamraju2020effect}. At mid-latitudes, the VER is typically higher in the summer hemisphere than in the winter hemisphere during the solstice, resulting in a large annual oscillation at mid-latitudes and a weak semiannual oscillation at the equator \cite{zhang2005response}. The observed seasonal variation primarily corresponded to the changes in the concentration of atomic oxygen in the MLT, which is influenced by solar activity and atmospheric dynamics such as temperature gradients and circulation patterns, resulting in changes in the distribution and transport of atomic oxygen.
The seasonal and latitudinal variation of daytime atomic oxygen green line emission has been investigated using UARS/WINDII \cite{gault1992wind} space-based observations from 1991-1997 \cite{zhang2005response} and ground-based measurements \cite{laskar2013investigations,laskar2015gravity, karan2016electrodynamic, karan2018effect, pallamraju2020effect}. Some of these studies reported that for the equatorial region, the variation has maxima during equinoxes. For mid-latitude in the northern and southern hemispheres, the green line emission has maxima during autumn and spring, respectively. 
Various other short and non-periodic variations modulated by gravity waves and tides in oxygen green line emissions are also reported \cite{shepherd2006airglow, liu2008seasonal}. 
Still, the variability of atomic oxygen green line emission with respect to various factors such as geomagnetic activity and solar cycle is not completely understood. Several studies in the past about airglow emission features and their implications on atmospheric dynamics have largely been event-based. Thus, the fundamental nature of the interrelation between various geophysical parameters and the concerned MLT variability has remained elusive. There is a strong need to look at various types of measurements and comprehensively understand the overall and long-term variability of the airglow emission, which can give us many valuable insights into the sun-earth interaction and energetics of the MLT region.

Understanding all the factors influencing the green line emission and their impacts on the VER is essential for interpreting observations and developing accurate upper atmosphere models. 
Various studies have been reported on the development of photochemical models to estimate the volume emission rate and intensity of green line airglow emission \cite{solomon2017global,krishna2009testing}; however, there are various uncertainties related to the variability of different parameters such as solar EUV, XUV, and UV fluxes during the solar cycle, ion and neutral densities have limited the model performance. Due to these limitations, the actual variability of green line emission due to various geophysical effects cannot be fully captured. Also, it is important to note that the airglow emission mechanism is altitude-dependent. All these factors limit the application of such photochemical models.

The availability of ground-based measurements of green line emission at 557.7 nm provides valuable insights into upper atmospheric processes and serves as an essential tool for remotely sensing atmospheric dynamics. Ground-based instruments, such as photometers or spectrometers, are utilized to detect and measure this specific emission. These instruments are strategically positioned at various locations worldwide to monitor the atomic oxygen green line emission. They provide essential data for space weather forecasting, atmospheric research, and validating theoretical models and significantly enhance our understanding of Earth's atmospheric dynamics and their implications.

However, these measurements are very limited and not widespread and are not as frequently available as desired. Also, the intensity measurements by photometers alone might not provide sufficient data to understand the vertical layered structure of the atmosphere comprehensively. Interpreting intensity-based measurements to retrieve vertical profiles often relies on assumptions. These assumptions might oversimplify the complex atmospheric interactions, leading to inaccuracies in determining vertical profiles. 
So, the limited availability of airglow measurements drives us to explore novel methodologies by utilizing the existing data more effectively rather than viewing the limited availability of airglow measurements as a barrier. By employing advanced techniques and using existing data along with complementary sources of information, we can develop robust models that offer insights into airglow behavior even with sparse measurements. The machine learning (ML) technique is one of the most advanced techniques that can be utilized to model the green line emission.

Machine learning-based modeling can potentially complement traditional physics-based models and make more realistic predictions to match the observations. This study uses a supervised machine learning technique to develop a predictive model for the volume emission rate of atomic oxygen green line emission. This model can predict VER for any given latitude, longitude, altitude, local solar time, and geophysical parameters. This model can be a powerful tool to predict the space weather fluctuations in the MLT region using the green line emission intensity as a precursor.

Given the highly complex nature of the interaction between the incoming energy and the upper atmosphere and the large volumes of measurements, it is extremely difficult to develop a comprehensive understanding of the MLT region. The decadal measurements of airglow by satellite-based instruments have provided high-quality data. With an appropriate choice of ML algorithm by carefully implementing the potential biases, we have attempted to develop a predictive model for the airglow emission rates. The ability of machine learning models to handle multidimensional and multivariate data can be particularly helpful in establishing subtle connections between several key parameters and measurable quantities in the upper atmosphere.

Developing a predictive machine learning model for the green line emission in Earth's upper atmosphere presents an exciting opportunity with certain challenges. This model can provide a more comprehensive understanding of green line emission at 557.7 nm, which is often sporadic and influenced by various factors. Additionally, accurately measuring several atmospheric parameters that impact this emission poses difficulties. Despite these obstacles, the innovative application of machine learning in this field offers a pathway to overcome these challenges.  The advances in machine learning methodologies and the development of new algorithms are promising for understanding the larger picture of space weather's influence on the geospace. The training data includes ten years of continuous observations of greenline emissions covering a multitude of latitudes, longitudes, and solar conditions. The conditions of training can be assumed to be a near superset of all possible variations in key variables. However, the model's prediction ability is only up to the connections it can develop within the limited dataset. It is important to note that the model can give a better estimate based on the climatological values. The limitation of predicting extreme variations which have not been witnessed will still remain. This paper presents an ML model developed to estimate and predict the vertical profile of atomic oxygen green line VER at 557.7 nm primarily based on WINDII measurements.

\section{Data and model description:}
\subsection{WINDII Observations}
The primary objective of the Upper Atmosphere Research Satellite (UARS), a NASA mission launched in 1991, was to investigate the various aspects of Earth's upper atmosphere and its interactions with the Sun. The Wind Imaging Interferometer (WINDII) was an instrument on board the UARS designed to measure atmospheric winds and temperatures \cite{shepherd1993windii}. 
The green line (557.7 nm) emission by atomic oxygen (O) is one of the key measurements made by WINDII. It measures the green line emission rate at 557.7 nm emitted by the excited state of atomic oxygen (O). The instrument utilizes a Michelson Interferometer that measures the Doppler shift as a phase shift of the co-sinusoidal interferogram generated by a single airglow line.WINDII was designed to measure wind, temperature, and emission rates of various species in the mesosphere and thermosphere between 80 and 300 km with a vertical resolution of typically 2-3 km in the mesosphere and about 5 km in the lower thermosphere. WINDII observations cover the latitude between 40$^o$S and 72$^o$N and 72$^o$S and 40$^o$N, depending on the yaw cycle of the satellite, which alternates in about 36 days. 
WINDII provides thirteen years (1991-2003) of measured data for atomic oxygen green line VER at 557.7 nm. The measurements have been carried out for an altitude range of about 80 to 300 km, which includes day- as well as night-time profiles of green line emission \cite{shepherd2012wind}. The present study utilizes the WINDII measured data from 1991 to 2003 to develop a predictive model to predict and estimate the VER of atomic oxygen green line emission for an altitude range of 80 to 300 km.

\subsection{Data and Model descriptions}
This paper presents a machine learning model to predict the volume emission rate (VER) of the green line (557.7 nm) by atomic oxygen in the upper atmosphere of Earth. The model is primarily based on the WINDII data measured during the period of 1991-2003. The model is developed for the target variable, i.e., VER at 557.7 nm. The geomagnetic indices that influence VER are obtained from OMNIWeb. The selection of various input features for the ML algorithm is based on the current state-of-the-art knowledge about the physical phenomenon that can influence the observed VER by atomic oxygen. The measurement dataset represents the full spectrum of relations and correlations between various geophysical parameters, which are state functions of the upper atmosphere. In this study, we assumed that the data includes all underlying relations between various input features and machine learning is capable of extracting even a very small correlation from the data with respect to the target variable. 

To develop and evaluate the performance and accuracy of the model, a dataset from 1991 to 2000 is selected from the entire dataset (1991 - 2003), and this subset is considered as the main dataset. The main dataset is further split into two subsets known as training- and testing- datasets. The training subset includes 90\% of the main dataset, and the remaining 10\% is being used as a testing subset. The training subset is exposed to machine learning algorithms to extract possible correlations and cross-correlations between the input and target variables. The remaining part of the entire dataset from 2001-2003 was unknown to the model and used to evaluate the model performance.
The list of all features and their respective sources used in this study are listed in Table \ref{tab1}. It is quite possible that there may be some other parameters beyond the listed variables which can influence the variability of VER.  Due to the lack of measurement datasets and knowledge, such parameters are excluded from the feature list in the current study. Among the various features investigated and listed in Table \ref{tab1}, the parameter SZA and altitude strongly correlate with the target variable. The reported correlation of green line VER with these parameters is based on the entire dataset (1991-2003) and has a value of -0.44 and -0.41 for SZA and altitude, respectively.

\begin{table}[h!]
\caption{List of input parameters, their description, sources, and unit.}
\centering
\begin{tabular}{c c c c}
\hline
Input parameter & Description & Unit & Source\\
\hline
F10.7&	Solar radio flux at (10.7 cm)&	SFU (10$^{-22}$ W/m$^{2}$/Hz)& OMNIWeb\\
S\_spot&	Sunspot number&		& OMNIWeb\\
LT & Local time & Hour & WINDII\\
SZA	&Solar zenith angle&	Degree&	WINDII\\
latitude&	Latitude&	Degree	&WINDII\\
longitude&	Longitude&	Degree & WINDII\\
altitude & Altitude & Kilometer & WINDII\\
\hline
\end{tabular}
\label{tab1}
\end{table}

\section{Algorithm and implementation}
The present work has been carried out using regression algorithms, which are based on supervised machine learning techniques, a subset of machine learning (ML), and artificial intelligence (AI). This is a very useful and well-established tool for developing highly accurate machine-learning models. This work has been implemented using three ensemble ML algorithms: Decision tree, Extreme Gradient Boosting (XGBoost), and Random Forest. The Decision trees are individual models that split data into branches based on feature value to make predictions, but they can over-fit when used as a single. XGBoost is another ensemble method that builds decision trees sequentially. Each new tree corrects the errors of the previous ones, making it highly accurate and efficient. A Random Forest is an ensemble of decision trees primarily used to reduce over-fitting by training each tree on random data subsets and then aggregating their results. All these algorithms are tested for the same dataset (training dataset 1991-2000). The performance of all these machine learning algorithms is based on their R-value (known as
the coefficient of linear correlation, which has a value between -1 and 1. The R-values closer to -1 or 1 indicate a higher correlation or strong relationship.  When comparing different algorithms, a higher R-value means the predictions are more accurate) is shown in Table \ref{tab2}.
\begin{table*}[h]
\centering
\caption{R-value of various ML algorithms for training and test dataset.}
\begin{tabular}{c |c| c}
\hline
ML/AI Algorithm &   \multicolumn{2}{c}{R-value}  \\

    & Training  & Testing\\
    \hline
    Decision Tree & 0.99 & 0.93\\
    XGBoost &	0.99&	0.95\\
    Random Forest&	0.99&	0.98\\
    \hline
\end{tabular}
\label{tab2}
\end{table*}

All the above-mentioned work has been implemented using Python programming language \cite{van2014python}, Scikit-learn \cite{pedregosa2011scikit}, and Keras \cite{chollet2015keras} libraries.

\section{Model performance}
The primary objective of the work presented in this paper is to develop a machine learning model that can estimate and predict the atomic oxygen green line volume emission rate (VER) in the mesosphere and thermospheric regions of Earth’s atmosphere. To achieve this, 13 years of WINDII measurement data of atomic oxygen airglow emission profiles are merged with OMNIWeb data. The implementation and performance of various ML algorithms have been discussed in section 3 (Algorithm and implementation). The model training and testing have been carried out using the training and testing subsets. The selection of data for training and testing subsets was random and selected from the main dataset (1991-2000) so that there would be no bias towards the prevalent conditions.  The performance of the model was evaluated using a new dataset measured by WINDII during 2001-2003. This dataset is never exposed either for training or testing the model, so it is a completely new dataset for the model and best fit to test the model performance.

\begin{figure}[h!]
\noindent\includegraphics[width=\textwidth]{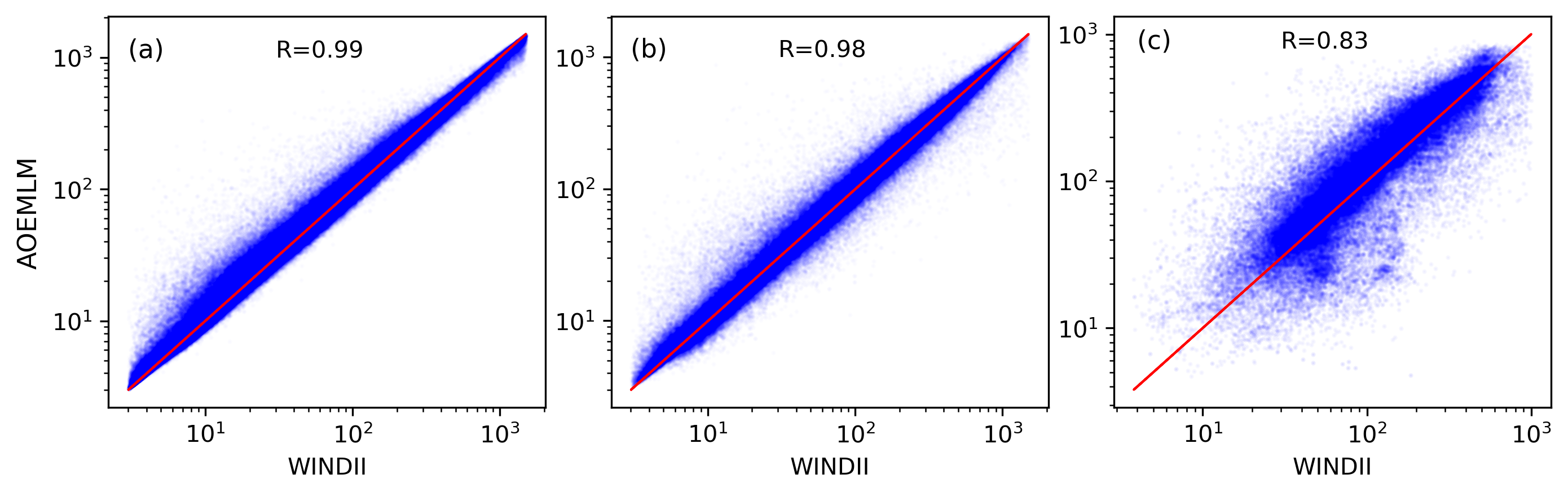}
\caption{The correlation between predicted (AOEMLM) and measured (WINDII) VER for (a) training, (b) test, and (c) for validation datasets.}
\label{fig2}
\end{figure}

All the algorithms have been trained and tested using the same dataset so that the accuracy and performance of models can be compared on the same database. Based on the higher R-value of algorithms, the random forest algorithm is selected for the further development of the model, which can make predictions that are compared with the actual measurements.
To evaluate the performance of the model, which has been trained as discussed above, the atomic oxygen green line VER is predicted by the model against the same input features (LT, SZA, F10.7, S\_spot, latitude, longitude, and altitude) for which WINDII measurements are available. The correlation between model-predicted VER and measured VER for both the dataset training and testing is shown in Figure \ref{fig2}.

\begin{figure}[h!]
\noindent\includegraphics[width=\textwidth]{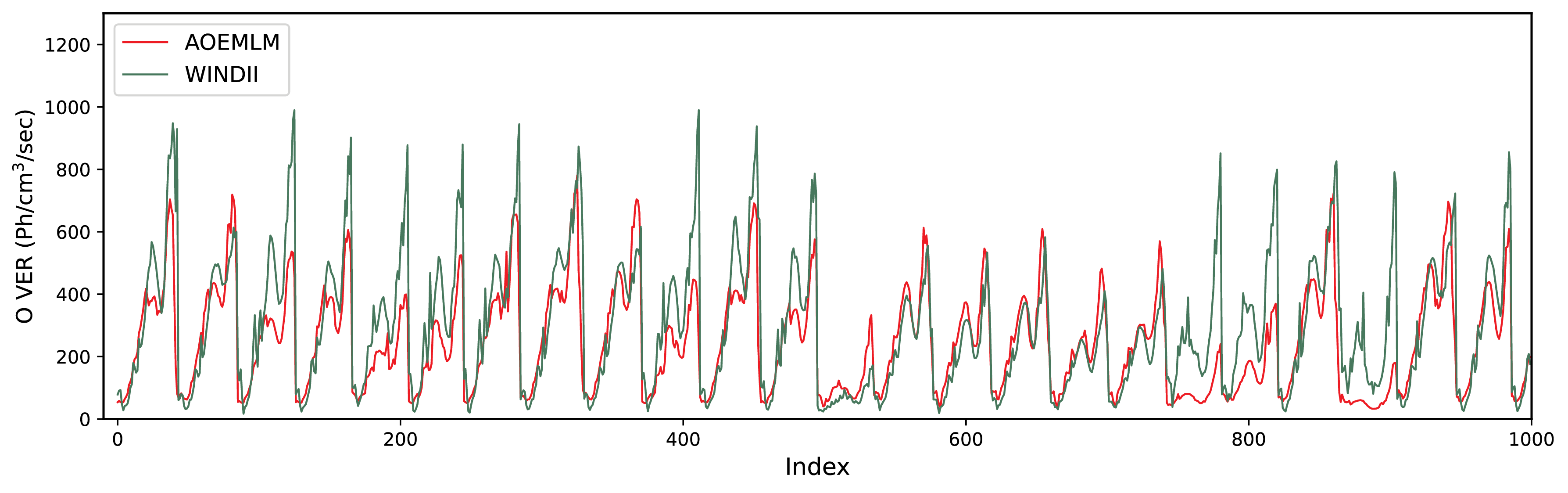}
\caption{The VER predicted by AOEMLM compared to the observations of WINDII. The samples were selected randomly from the dataset (2001-2003).}
\label{fig3}
\end{figure}

The red line in Figure \ref{fig2} is the best-fit line, having zero deviation between the predicted and measured values of volume emission rates. The deviation from the red line represents the difference between the measured value and model prediction. To quantify the model's accuracy, the R-value has been given for training, testing, and validation datasets. Figure 2 shows the correlation between AOEMLM's prediction and actual WINDII measured VER for the training, testing, and validation dataset. As shown in Figure 2, an R-value of 0.99 for the training, 0.98 for the test, and 0.83 for the validation datasets was achieved using the Random Forest algorithm. The higher value of correlations (closer to 1) indicates the better accuracy and trustworthiness of the model. Also, it shows the model's capability to predict green line VER based on the input features. The prediction capability of the model is compared with the WINDII measurements and the GLOW model (v.0.981) \cite{solomon2017global}. A comparison of the measured volume emission rate by WINDII between 2001-2003 and predicted VER by the AOEMLM for the same inputs is shown in Figure \ref{fig3}. 

\begin{figure}[h!]
\noindent\includegraphics[width=\textwidth]{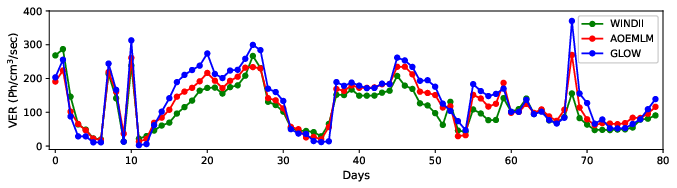}
\caption{The daily averaged VER predicted by AOEMLM compared to GLOW and WINDII for the validation dataset (2001-2003).}
\label{fig4}
\end{figure}

The term Index represents the set of input parameters (listed in Table 1) that correspond to the predicted green line VER. Each index value represents the predicted VER value corresponding to each set of input parameters and is compared with the WINDII measurements. A set of such individual input parameters corresponding to the same location and local time generates a green line VER profile, which can be seen in Figure \ref{fig3}. It is seen that the model predictions match the WINDII measurements. However, these few profiles alone are insufficient to demonstrate the robust prediction accuracy of AOEMLM. To further validate the results, we considered the entire validation dataset (2001-2003) and calculated the daily average green line VER for all measurements at all altitudes. Figure \ref{fig4} presents the comparison of the daily averaged green line VER estimated by AOEMLM, GLOW, and measured by the WINDII. From Figure \ref{fig4}, it can be seen that the daily averaged green line VER predicted by AOEMLM matches quite well with the actual measurements of WINDII. To quantify the performance of both the models, the mean deviation is computed, resulting in the values of 14.95 ph.cm$^{-3}$.s$^{-1}$ and 28.43 ph.cm$^{-3}$.s$^{-1}$ for the AOEMLM and GLOW models, respectively.

As discussed in earlier sections, machine learning models are able to find even very small correlations between features and make very fine connections, thus determining the relative weights of input features with respect to the target parameter. They are also able to handle and correlate the fluctuations in target parameters induced by the variability of various input features. As it is known, the ionospheric density shows strong seasonal, diurnal, and solar cycle variability. The airglow emission intensity strongly depends on the ionospheric densities. Thus, it is very important to evaluate the model performance for various seasons and compare it with similar measurements made by WINDII. The ability of this model to estimate and predict the atomic oxygen green line volume emission rate in different seasons of the year is shown in Figure \ref{fig5}, \ref{fig6}, and \ref{fig7}. The seasonal variation of VER and the ability of the ML model (AOEMLM) to capture the variation with latitude, solar zenith angle (SZA), different geomagnetic conditions, and the influence of the solar cycle along with local time are discussed in the upcoming sections.

\section{Variation of green line VER with solar zenith angle (SZA):}
The SZA is an important parameter that characterizes the solar energy input during the day as a function of time at any given latitude and longitude. It varies with time of day, latitude, and time of year (it has different values at different seasons of a year at the same place and local time). SZA is a very useful parameter to determine the amount of solar radiation that reaches the Earth's surface and can be used to understand the energy balance of the Earth's atmosphere. The SZA affects the amount of solar flux received by the upper atmosphere of Earth, which results in the production of atomic oxygen \cite{maharaj2004solar, zhang2005response}. The SZA indicates the position of the Sun at any given time of the day with respect to the normal.  For a smaller SZA, the Sun is positioned close to the overhead, resulting in higher solar radiation reaching the Earth's surface. Hence, the emission rate of the green line (557.7 nm) is enhanced due to the increased photo-ionization of atomic oxygen and the subsequent recombination process of O+ ions with electron or neutral atoms \cite{witasse1999modeling}.

As the SZA increases, the incident radiation has to travel more distance to reach a given airglow altitude; thus, due to the attenuation, the solar flux at the airglow emission altitude decreases. This results in lower ionization of atomic oxygen, leading to lower green line (557.7 nm) emissions. Along with SZA, the atomic oxygen green line emission is also influenced by several other factors, including concentration and reactant availability at a given altitude, latitude, and solar flux (F10.7). The combined effect of all these reactants makes it very complex and difficult to estimate the emission rate accurately. The strong dependence of green line emission on SZA makes it an important parameter for interpreting Earth's upper atmosphere observations.

The WINDII instrument, during its observational cycle, has provided a wealth of information on the variation of atomic oxygen green line emission in the upper atmosphere. The observations suggest that the emission rate is strongly dependent on the SZA. The WINDII measurements of the green atomic oxygen line (557.7 nm) at various latitudes and SZA show such variation, and two peaks centered around 100 km (E-layer) and 160 km (F-layer) were observed. The observations show that the lower peak height in the E-layer shows no apparent relation with SZA and latitudes; however, the upper peak in the F-layer shows upward shifts in altitude with increasing SZA. Also, the peak emission rate decreases with increasing the value of SZA. The observations also show that the 557.7 nm green line emission in the Earth's upper atmosphere also exhibits a strong seasonal variation. This variation is primarily driven by changes in the amount of solar radiation that reaches the upper atmosphere, which is influenced by the tilt angle of Earth and its orbit around the Sun \cite{zhang2005response}.

\begin{figure}[h!]
\begin{subfigure}
\noindent\includegraphics[width=\textwidth]{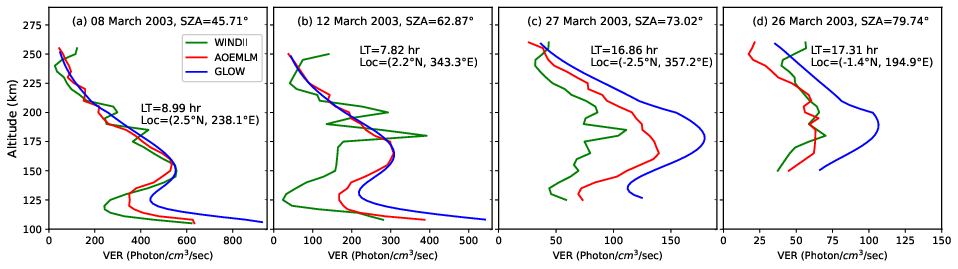}
\end{subfigure}
\begin{subfigure}
\noindent\includegraphics[width=\textwidth]{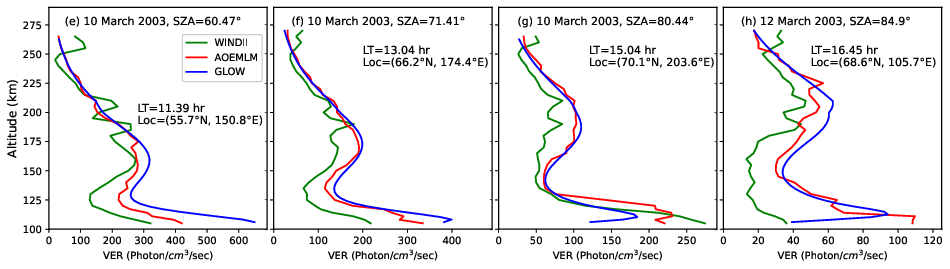}
\end{subfigure}
\vspace{-0.8cm}
\caption{Variation of O($^1$S) greenline volume emission rate with SZA at lower (a-d) and higher (e-h) latitudes during the March equinox season.}
\label{fig5}
\end{figure}

\begin{figure}[h!]
\begin{subfigure}
\noindent\includegraphics[width=\textwidth]{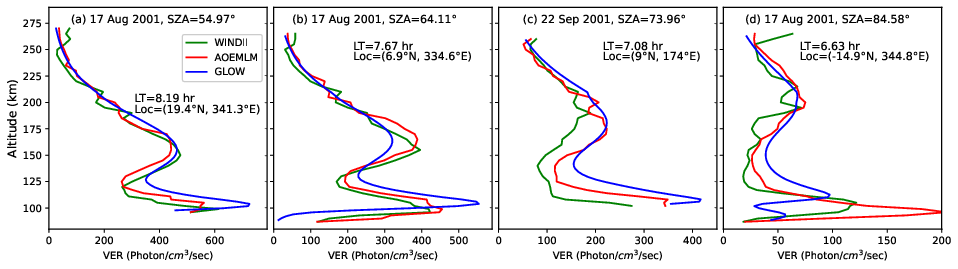}
\end{subfigure}
\begin{subfigure}
\noindent\includegraphics[width=\textwidth]{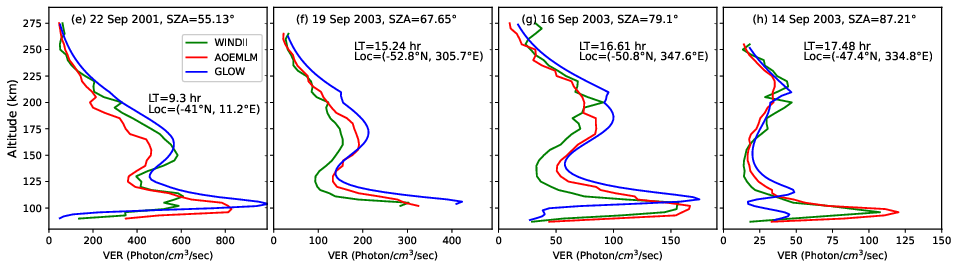}
\end{subfigure}
\vspace{-0.8cm}
\caption{Variation of O($^1$S) greenline volume emission rate with SZA at lower (a-d) and mid-higher (e-h) latitudes during the September equinox season.}
\label{fig6}
\end{figure}

\begin{figure}[h!]
\begin{subfigure}
\noindent\includegraphics[width=\textwidth]{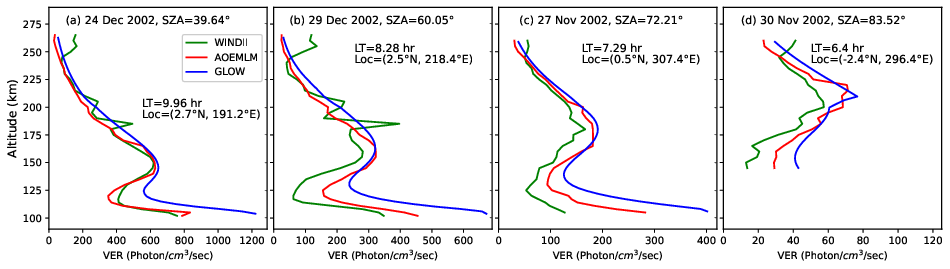}
\end{subfigure}
\begin{subfigure}
\noindent\includegraphics[width=\textwidth]{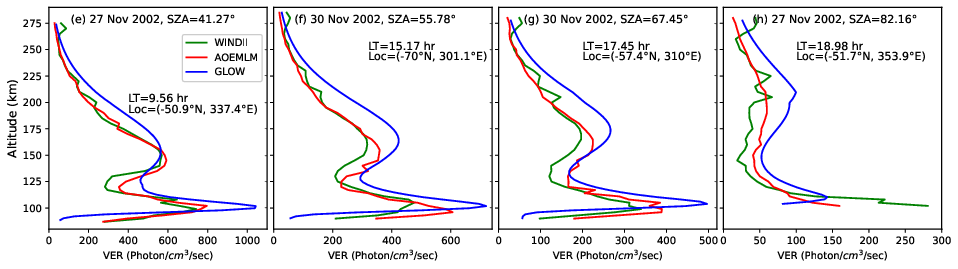}
\end{subfigure}
\vspace{-0.8cm}
\caption{Variation of O($^1$S) greenline volume emission rate with SZA at lower (a-d) and mid-higher (e-h) latitudes during December solstice season}
\label{fig7}
\end{figure}

The ability of AOEMLM to capture the variation in VER with SZA and season is tested using the WINDII measured profiles and compared with the Global Airglow (GLOW) model \cite{solomon2017global}. All the profiles are selected from the validation dataset, which was never used earlier in the model development. The results of the comparison between the AOEMLM predictions and satellite and GLOW model at various latitudes and SZA during the March equinox are shown in Figure \ref{fig5}. The idea is to test the model for a wide range of solar zenith angles and its accuracy for the observed variability in volume emission rate, as discussed above. It is clearly visible from Figure \ref{fig5} that with the increase in SZA, the F-layer peak shifts towards higher altitudes, and the value of peak emission volume emission rate decreases. A similar pattern is also observed during the September equinox and December solstice, as shown in Figures  \ref{fig6} and  \ref{fig7}. AOEMLM captures all these variations very well, and the predictions match quite well with the measured values. 

It is evident from Figures \ref{fig5},  \ref{fig6} \&  \ref{fig7} that the AOEMLM prediction shows both the variation and shift in peak altitude, which are the main features of the observed variability. The predictions of AOEMLM are closer to the actual measurements of WINDII and compared with the existing GLOW model results. This shows that AOEMLM is capable of extracting the information from the input feature and estimating the target (VER). For latitudinal variation, there was no clear variability observed. Overall, the latitudinal variation of green line emission is a complex and dynamic phenomenon that is still a subject of ongoing research and investigation. However, the model successfully predicts the vertical profiles of green line VER with very good accuracy. All these results show that AOEMLM can be used globally to calculate and predict the green line volume emission rate for any geophysical conditions.

\section{Variation of O($^1$S) green line VER with F10.7 Solar Flux}
\begin{figure}[h!]
\begin{subfigure}
\noindent\includegraphics[width=0.5\textwidth]{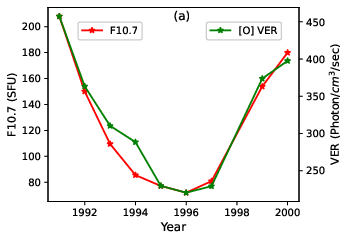}
\end{subfigure}
\begin{subfigure}
\noindent\includegraphics[width=0.5\textwidth]{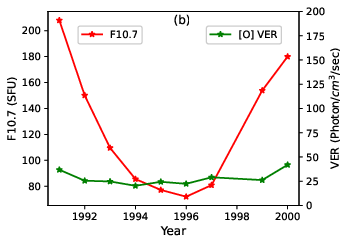}
\end{subfigure}
\vspace{-0.8cm}
\caption{The variation of (a) daytime and (b) nighttime annual averaged green line VER with F10.7 flux during the 22-23 solar cycle (1991-2000).}
\label{fig8}
\end{figure}

Solar flux F10.7, also known as the 10.7 cm radio flux or the F10.7 index, is a measure of the radio emissions at a wavelength of 10.7 centimeters from the chromosphere of the sun. It is used as a proxy for solar activity and the strength of the solar cycle \cite{tapping2017changing,le2012secular}. It is an important parameter for space weather forecasting, as it is directly related to the degree of ionization in the Earth's ionosphere \cite{balan1994variations}. From the WINDII observation, it is observed that there is a very strong correlation between the intensity of the atomic oxygen green line (557.7 nm) and the solar flux F10.7. During high solar activity, the increased ionizing radiation from the Sun enhances the green line emission. Conversely, during periods of low solar activity, the electron density decreases, resulting in weaker green line emission. Figure \ref{fig8} shows the daytime (0 $<$ SZA $<$ 40 degrees, the part facing to the Sun ) and nighttime (140 $<$ SZA $<$ 180 degrees, the part opposite to the Sun ) annual average green line VER (calculated for each day of the year including the entire range of altitudes) with the solar flux (F10.7) for solar cycle 22-23 (1991-2000) ).

\begin{figure}[h!]
\centering
\begin{subfigure}
\noindent\includegraphics[width=\textwidth]{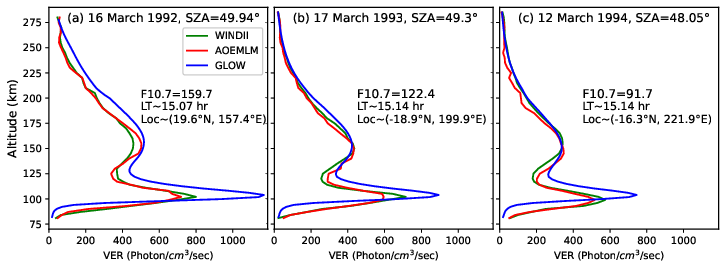}
\end{subfigure}
\begin{subfigure}
\noindent\includegraphics[width=0.67\textwidth]{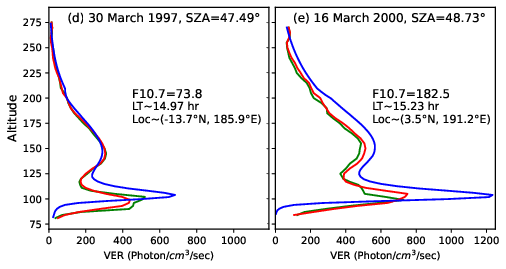}
\end{subfigure}
\caption{Variation of VER with solar flux (F10.7) for 22nd solar cycle  at lower latitudes}
\label{fig9}
\end{figure}

\begin{figure}[h!]
\noindent\includegraphics[width=\textwidth]{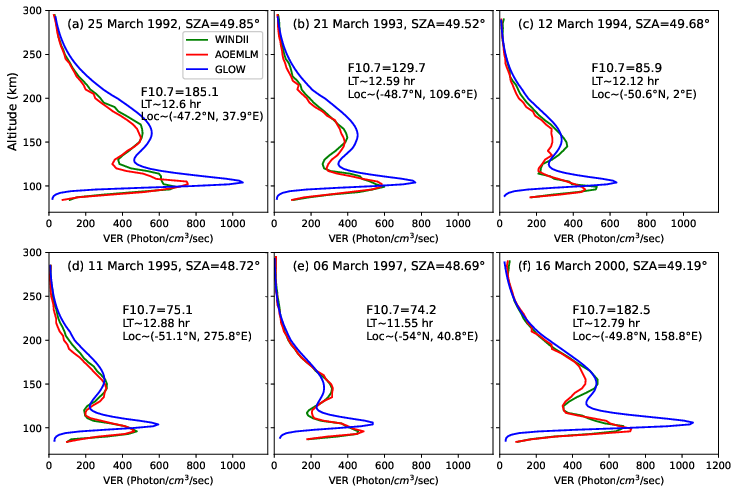}
\caption{Variation of VER with solar flux (F10.7) for 22nd solar cycle  at mid-latitudes.}
\label{fig10}
\end{figure}

\begin{figure}[h!]
\begin{subfigure}
\noindent\includegraphics[width=0.5\textwidth]{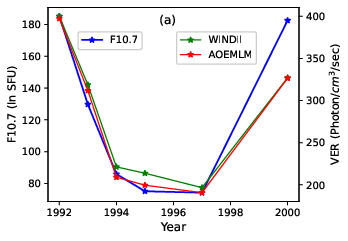}
\end{subfigure}
\begin{subfigure}
\noindent\includegraphics[width=0.5\textwidth]{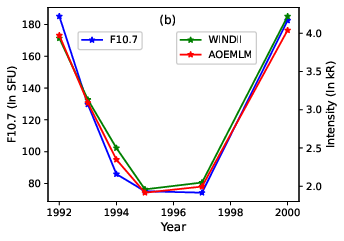}
\end{subfigure}
\vspace{-0.8cm}
\caption{The variation of daytime (a) daily averaged VER and (b) intensity with solar flux (F10.7).}
\label{fig11}
\end{figure}

The relationship between the atomic oxygen green line emission and solar flux F10.7 is affected by several factors, such as the latitude and altitude of observation, as well as the presence of other atmospheric constituents \cite{bag2017study, zhang2005response}. However, the 10-year WINDII measured atomic oxygen green line emission training data includes all such variability, and ML can extract the information from input features, as discussed earlier. So, AOEMLM is able to predict all variability with respect to solar flux F10.7. The variability of VER with F10.7 and solar cycle 22 for a few profiles for lower (0$<$latitude$<$30$^o$N or S) and mid-latitudes (30$<$latitude$<$60$^o$N or S) are shown in Figure  \ref{fig9} and \ref{fig10}, respectively. The results of the AOEMLM are also compared with the existing GLOW model to make the AOEMLM reliable. Figure  \ref{fig11} shows the variation of daytime daily averaged VER and intensities with solar flux (f10.7) corresponding to the days shown in Figure \ref{fig10}.

As seen from Figures \ref{fig9} and \ref{fig10}, the AOEMLM predictions have very good agreement with the WINDII measurements. The AOEMLM has good accuracy in predicting variation in VER, peak VER, and peak emission height with the variation of the solar cycle or F10.7 flux. The results are also compared with the existing GLOW mode. From Figures \ref{fig11}, it is seen that the AOEMLM predictions also match well for the daily averaged green line VER and the intensity.

\vspace{-0.8cm}

\vspace{0.5cm}
\section{Predictions of AOEMLM in comparison to MIGHTI observation}
\begin{figure}[h!]
\begin{subfigure}
\noindent\includegraphics[width=\textwidth]{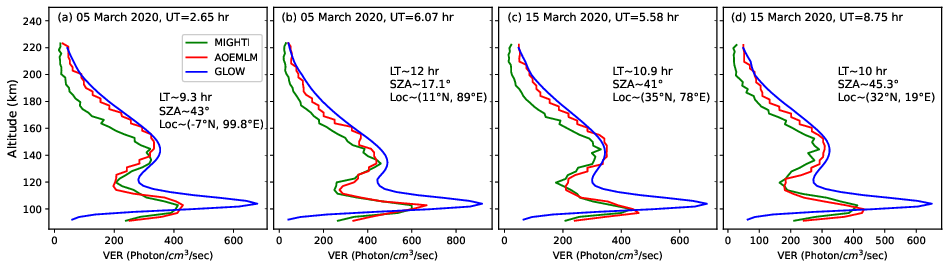}
\end{subfigure}
\begin{subfigure}
\noindent\includegraphics[width=\textwidth]{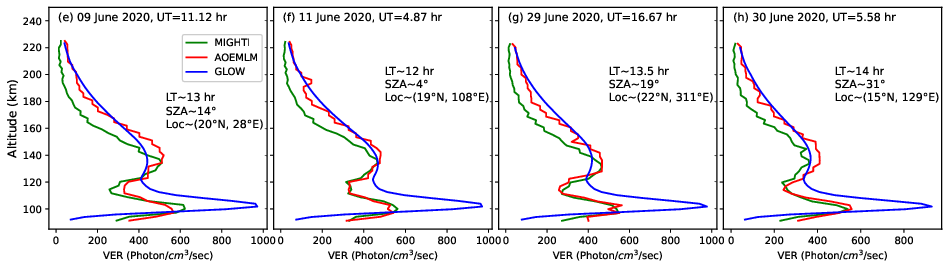}
\end{subfigure}
\begin{subfigure}
\noindent\includegraphics[width=\textwidth]{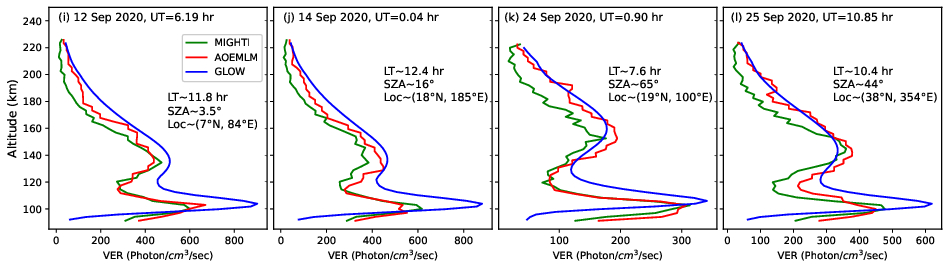}
\end{subfigure}
\begin{subfigure}
\noindent\includegraphics[width=\textwidth]{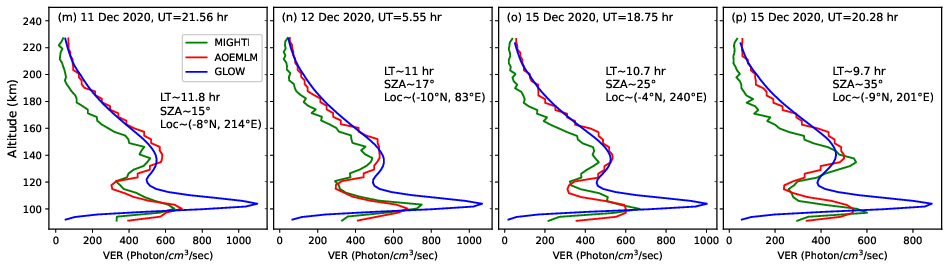}
\end{subfigure}
\vspace{-0.8cm}
\caption{The seasonal variation of green line VER during March (a-d), September (i-l) equinox and  June (e-h), December (m-p) solstice season.}
\label{fig12}
\end{figure}

The Michelson Interferometer for Global High-resolution Thermospheric Imaging (MIGHTI) instrument was launched on 10 October 2019 onboard NASA’s Ionospheric Connection Explorer (ICON) satellite. It was designed to measure thermospheric horizontal wind velocity and temperature profiles in the altitude range from 90 km to 300 km. For the wind measurements, it uses the Doppler shift of the atomic oxygen green- (557.5 nm) and red- (630 nm) line emissions measured by two perpendicular fields of view pointed at the Earth’s limb \cite{englert2017michelson}. MIGHTI provides the atomic oxygen green and red line emission data from 6 December 2019 onward. To test the validity of AOEMLM for the long term (about 20 years), we compared the AOEMLM and GLOW models predicted atomic oxygen VER (557.7 nm) with the observations of MIGHTI during the solstices and equinox seasons. The comparison of the results is shown in Figure \ref{fig12}.

\begin{figure}[!h]
\noindent\includegraphics[width=\textwidth]{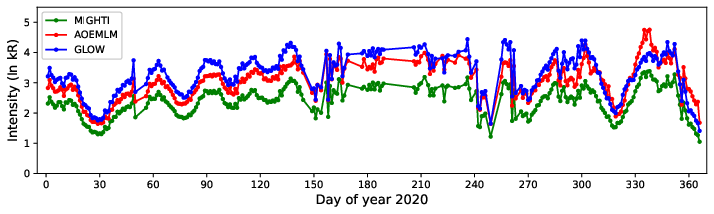}
\vspace{-0.9cm}
\caption{The daily averaged atomic oxygen green line intensity predicted by AOEMLM compared to MIGHTI observation for the year 2020.}
\label{fig13}
\end{figure}

It is seen from Figure \ref{fig12} that the AOEMLM predictions of vertical profiles of green line VER have better consistency with MIGHTI observation and good accuracy in predicting the VER in the E-region of the ionosphere. However, a noticeable deviation from MIGHTI observations is observed in the F-region of the ionosphere. 
Additionally, to test the AOEMLM on a larger dataset, we calculated the daily average intensity of green line emissions by integrating the vertical profiles for the MIGHTI dataset of the year 2020 for each day and compared across AOEMLM, GLOW, and MIGHTI observations as shown in Figure \ref{fig13}. The mean deviation was also calculated for the AOEMLM and GLOW models to evaluate model performance, yielding the values of 0.66 kR and 0.88 kR, respectively. The model AOEMLM was trained and developed using the WINDII observation from 1991 to 2003. However, the 20-year-later predictions of the AOEMLM are close to the actual observations of MIGHTI with consistent accuracy with the season (solstice and equinox), as shown in Figure \ref{fig12} \& \ref{fig13}.
These results show that the AOEMLM is a reliable ML model and can be used to predict the green line VER under varying geophysical conditions.

\vspace{-1.5cm}

\vspace{5cm}
\section{Comparison of AOEMLM with ground-based measurements}
So far, we have carried out a comparison of AOEMLM with satellite-based measurements from WINDII and MIGHTI, which use limb scanning to observe the atmosphere. While observing the green line emission from the ground, one obtained a column-integrated emission wherein both lower and higher altitude peaks of the green line emission are merged and provide a single value, which is measured by a photometer from a given location.  This method allows us to study the temporal variation in total emission. To evaluate the performance of AOEMLM in capturing the temporal variations in the daytime atomic oxygen green line emission at 557.7 nm, we compared the model results with the ground-based green line dayglow daytime measurements obtained from (17.5°N, 78.4°E), India using the observations of Multiwavelength Imaging Spectrograph using Echelle grating (MISE) \cite{pallamraju2013mise}.
\begin{figure}[h!]
\noindent\includegraphics[width=\textwidth]{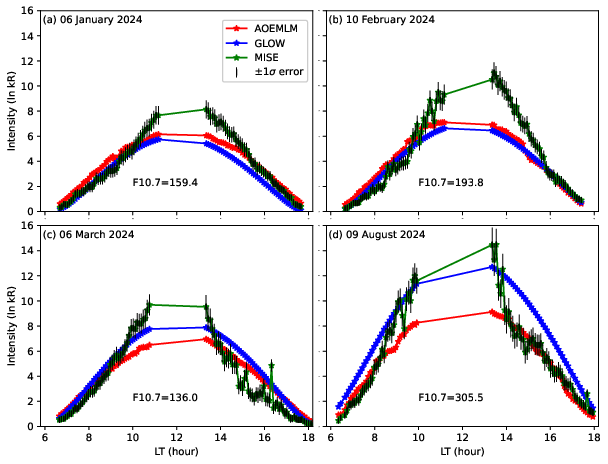}
\vspace{-0.8cm}
\caption{The variation of daytime atomic oxygen green line intensity with local time as measured by MISE along with $\pm1\sigma$ uncertainties and estimated by AOEMLM and GLOW model are shown.}
\label{fig14}
\end{figure}

A sample of four days obtained in different months in 2024 is considered for comparison. The AOEMLM and GLOW models were used to estimate the vertical profile of the green line VER corresponding to the observational latitude. These profiles were then integrated with respect to altitude to calculate the total green line emission intensity, enabling a fair comparison with the ground-based photometer measurements. It is important to note that MSIE captures entire green line emissions originating across the entire range of altitudes. However, the AOEMLM and GLOW models are altitude-dependent and were integrated over the column in the 90 to 220 km altitude range. This range includes the primary region where a major part of the green line emissions occurs, ensuring a meaningful comparison. Figure \ref{fig14} presents the green line intensity measured by the MSIE and its comparison with the predicted values of the AOEMLM and GLOW models. The results demonstrate that the AOEMLM predictions align closely with the ground-based data, showing the capability of AOEMLM to capture the variability of green line airglow intensity. MISE is a slit spectrograph. Around the local noon, sunlight directly enters the slit and saturates the detector, resulting in a data gap during this time.

It can also be seen from Figure \ref{fig14} that around noon, when the sun is nearly overhead, the MISE records higher green line intensities, which can be larger in either pre-noon or post-noon. For example, the post-noon enhancement is seen on 06 January and 10 February, whereas on 06 March and 09 August, diurnal emission variability shows a pre-noon peak. This asymmetric behavior in the ground-based measurements has been demonstrated earlier  \cite{karan2016electrodynamic} to be dependent on the equatorial electrodynamical behavior, which changes from day to day. Therefore, non-physics-based methods, such as the AOEMLM, are not expected to predict such behavior accurately, and so during this time period, both the models AOEMLM and GLOW show a higher deviation from the actual measurements. Other times, AOEMLM shows a reasonably good agreement with the MISE measurements.

\section{Discussion and Summary}
This study presents a machine learning modeling approach to predict the vertical volume emission rate profiles of the atomic oxygen green line (557.7 nm) in the upper atmosphere. This study provides a new modeling technique by utilizing solar and geomagnetic parameters. The machine learning model (AOEMLM) is developed using the measured data of WINDII using the supervised ML algorithm known as random forest. For the development of AOEMLM, all primary solar and geomagnetic parameters are considered, which are representative of various drivers having the potential to influence the chemistry and energy balance in the upper atmosphere. The target parameter of this study is the atomic oxygen green line volume emission rate. The training and testing of AOEMLM have been done using the WINDII dataset and Random Forest ML algorithm, as mentioned in Section 2 (data and model description) \& section 3 (Algorithm and implementation). It has been consistently observed that AOEMLM has very good training, testing, and validation scores of 0.99, 0.98, and 0.83, respectively. The model is also validated using the actual WINDII measurements from 2001-2003 and compared with the GLOW model estimation. It is important to note that the data from 2001-2003 was never used for either training or testing. The AOEMLM model has been used to predict atomic oxygen green line VER for various seasons (solstice and equinox), latitudes, SZA, and solar flux (F10.7). The dependency and variation of green line (557.7 nm) emission on SZA and F10.7 are well captured by AOEMLM.
These results show that the AOEMLM model successfully captures all the variability in atomic oxygen green line vertical profiles with very good accuracy. Furthermore, the AOEMLM’s predictions are verified using the MIGHTI and photometric measurements obtained from MISE and compared with the existing Global Airglow model (GLOW). The comparisons with MIGHTI, ground-based measurements, and the existing chemistry-based GLOW model show that the ML approach is capable of handling the complex phenomena occurring in the upper atmosphere of the Earth and can be used to predict and forecast the influence of space weather.
Based on the results, it can be stated that machine learning is capable of extracting the fine relations between inputs and is able to estimate and predict the target with good accuracy, and performance can be enhanced using a higher amount of data. This study provides a very useful and effective tool to predict the airglow emission rate, which is an important tracer for understanding the physical, chemical, and dynamic state of the upper atmosphere. The AOEMLM model can be a valuable asset for aeronomy research.

\section{Data Availability Statement}
The results of this study can be regenerated using the Python files available on the GitHub repository using the link (\url{https://github.com/Dnailwal/AOEMLM.git}).
The source datasets can be accessed through the following data providers:

The WINDII data is publicly available through the open data and information Portal of the Canadian Space Agency (\url{https://donnees-data.asc-csa.gc.ca/users/}), MIGHTI data can be downloaded from the ICON-MIGHT data portal using FTP web browsers (\url{ftp://icon-science.ssl.berkeley.edu/pub/}), and the OMNIWeb data can be accessed using the link (\url{https://omniweb.gsfc.nasa.gov/form/dx1.html}) from the NASA's space physics data facility (SPDF). The GLOW profiles were generated using Python libraries from the NCAR-GLOW GitHub repository (\url{https://github.com/space-physics/NCAR-GLOW}). The green line emission dayglow data used in this work were obtained by the Physical Research Laboratory and can be made available on request.

\section{Acknowledgments}
We acknowledge the WINDII, OMNIWeb, MIGHTI, and Global Airglow model (GLOW) for providing all the necessary data for developing this atomic oxygen emission model AOEMLM. The daytime ground-based airglow experiment was carried out by the Physical Research Laboratory (PRL), Ahmedabad. The efforts of DP at PRL are supported by the Department of Space, Government of India. One of the authors, D. Nailwal, thanks the Ministry of Education (MoE), Government of India, for the financial support as a graduate assistantship.

\bibliography{bibliography.bib} 

\end{document}